\pdfoutput=1
\documentclass[aps,prl,showpacs,floats,twocolumn,amsfonts,amssymb,unsortedaddress,floatfix]{revtex4}
\input epsf
\usepackage{bm}
\usepackage{amsmath}
\usepackage{graphicx}

\begin{document}

\newcommand{\ud}{\mathrm{d}}
\newcommand{\eq}{\begin{equation}}
\newcommand{\en}{\end{equation}}
\newcommand{\eqa}{\begin{eqnarray}}
\newcommand{\ena}{\end{eqnarray}}
\newcommand{\eqan}{\begin{eqnarray*}}
\newcommand{\enan}{\end{eqnarray*}}

\newcommand{\Dslash}{{\slash{\kern -0.5em}\partial}}
\newcommand{\Aslash}{{\slash{\kern -0.5em}A}}

\def\d{\partial}
\def\({\left(}
\def\){\right)}

\title{Spiraling Solitons: a Continuum Model for Dynamical Phyllotaxis and Beyond}
\author{Cristiano Nisoli}
\affiliation{Theoretical Division and Center for Nonlinear Studies, Los Alamos National Laboratory, Los Alamos NM 87545 USA}

\date{\today}
\begin{abstract}
A novel, protean, topological soliton has recently been shown to emerge in systems of repulsive particles in cylindrical geometries, whose statics is described by the number-theoretical objects of phyllotaxis. Here we present  a minimal and local continuum model that can explain many of the features of the phyllotactic soliton, such as locked speed, screw shift, energy transport and,  for Wigner crystal on a nanotube, charge transport. The treatment is general and should apply  to other spiraling systems. Unlike e.g. Sine-Gornon-like systems, our solitons can exist between non-degenerate structure, imply a power flow through the system, dynamics of the domains it separates; we also predict  pulses, both static and dynamic.  Applications include charge transport in Wigner Crystals on nanotubes or A-  to B-DNA transitions. 
\end{abstract}



\maketitle

\section{Introduction.}
The topological soliton, the moving domain wall between degenerate structures, ubiquitously populates systems of discrete symmetry, most notably the Ising model, and appears at different scales and in many realms of physics~\cite{solitons}: in mechanical  or electrical apparati~\cite{solitons,Scott69, Nakajima}, superconducting Josephson junctions~\cite{Ustinov}, non-perturbative theory of quantum tunneling~\cite{Kleinert} and particle physics (e.g. Yang-Mills monopoles and instantons~\cite{Actor}, sigma model lumps and Skyrmions~\cite{solitons2}).

Recently Nisoli {\it et al.}~\cite{Nisoli} have discovered a novel kind of  topological soliton in systems of repulsive particles whose degenerate statics is dictated by the intricate and fascinating number-theoretical laws of phyllotaxis. As the phenomena they describe are purely geometrical in origin, one would expect this kind of ``phillotactic'' soliton to play a role in many different physical systems at different scales, wherever repulsive particles in cylindrical geometries are present~\cite{Nisoli}. 

Physically, the problem appears to he highly non local, as energy and momentum is not confined inside the soliton but flows though it between the rotating domains it separates (see below). Yet, in this article we show that a minimal, local, continuum model for the phyllotactic soliton that subsumes  the effect of the kinetic energy of the domains into a modification of the potential energy in the  familiar equivalent newtonian problem~\cite{Kleinert}, can correctly predict its speed, the  transfer of energy between boundaries, the observed shift in screw angle, and charge/density variations, as well as classify its rather complicated zoology. 
\section{From Levitov's to dynamical phyllotaxis.}
\begin{figure}[t!!]
\center
\hspace{10 mm}\includegraphics[width=2.4 in]{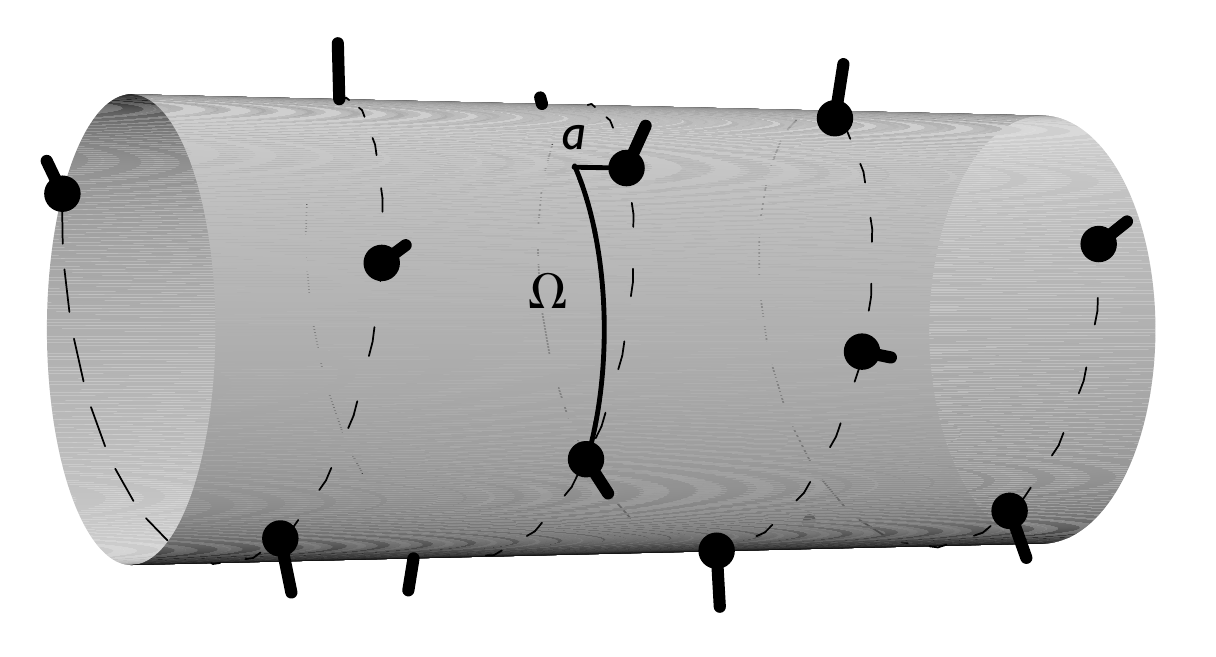}\vspace{0 mm}

\includegraphics[width=2.9 in]{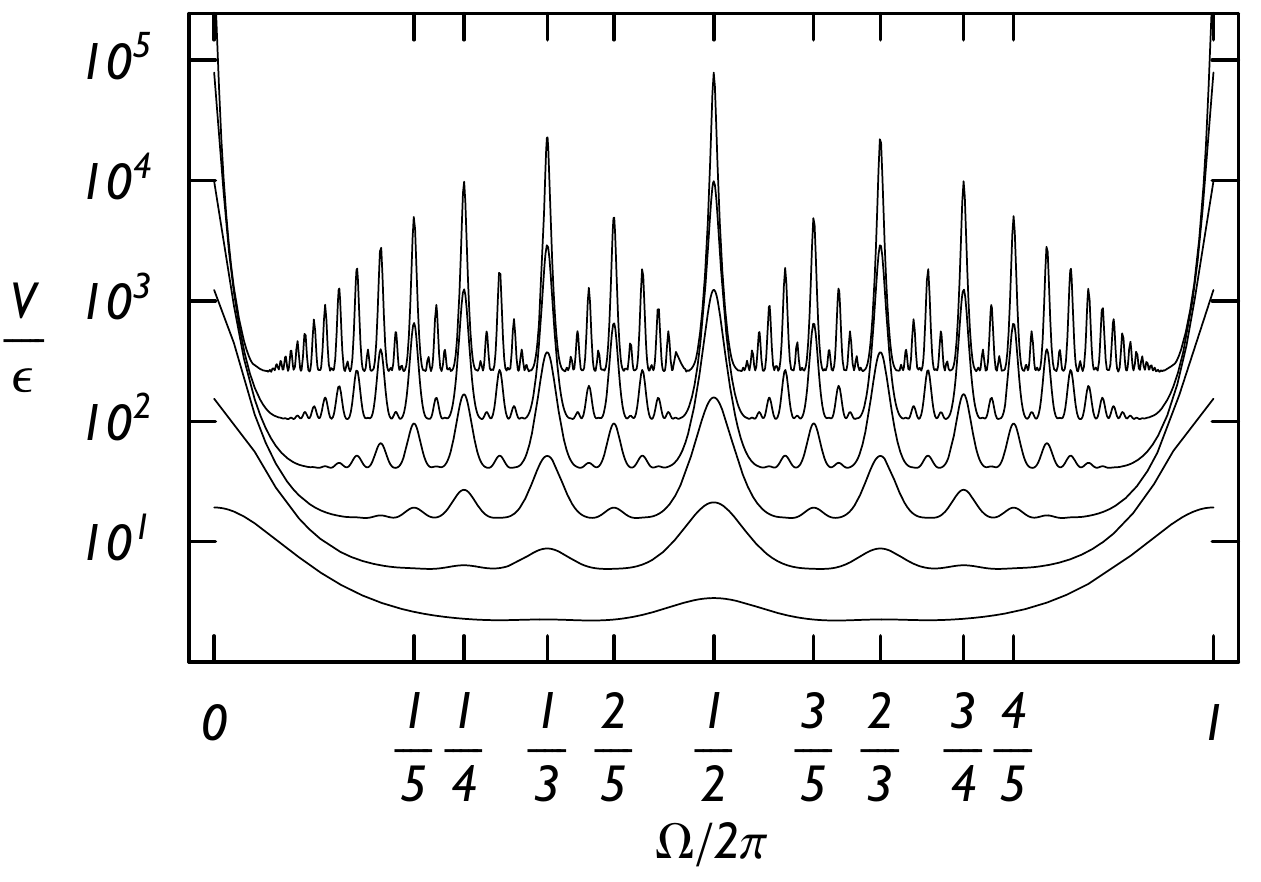}
\caption{ Top: the dynamical cactus, repulsive particles, here dipoles, on the surface of a cylinder.  $a$ is the distance between consecutive rings, $\Omega$ and their angular shift.  The dashed line is the so-called  ``generative spiral''~\cite{his1}. Bottom: for a dipole-dipole interaction, energy $V$ versus screw angle $\Omega$ (semi-logarithmic plot) for halving values of the parameter $a/R$ starting from $0.5$ (lowest line), in arbitrary units $\epsilon$.}
\label{fig1}
\end{figure}
In the early nineties, while studying vortices in layered superconductors, Levitov~\cite{Levitov} showed that a system of repulsive particles on a long cylinder presents a rich degenerate set of ground states; these can be labeled by number-theoretical objects, the Farey classes, that arise in phyllotaxis, the study of dispositions of leaves on a stem and other self-organized structures in botany~\cite{his1}. While it is still debated whether Levitov's model is the best suited to explain botanical phyllotaxis, it  certainly imported the same mathematical structures to describe  a variety of physical systems. Already phyllotactic patterns had been seen or predicted outside the domain of botany~\cite{his2}, in polypeptide chains~\cite{polypeptide},  cells of Benard convection~\cite{Benard}, or  vortex lattices in superconductors~\cite{Levitov}. 

\subsection{Dynamical Cactus}

Recently Nisoli {\it et al.}~\cite{Nisoli} have shown numerically and experimentally that physical systems whose statics is dictated by phyllotaxis can access dynamics in ways unknown in botany. 
 They have investigated the statics and  dynamics of a prototypical phyllotactic system, the dynamical cactus of Figure~\ref{fig1}, and found that while its statics faithfully reproduce phyllotaxis, its dynamics reveal new physics beyond pattern formation. 
In particular, the set of linear excitations contains classical  rotons and maxons; in the nonlinear regime, both dynamical simulations and experiments revealed a novel highly structured family of topological solitons that  can change identity upon collision, and posses a very rich dynamics~\cite{Nisoli}. 
Unlike other topological solitons, the phyllotactic one separates domains of different dynamics; these domains can store kinetic energy while rotating around the axis of the dynamical cactus; energy and momentum flow across the soliton. Moreover as those domains are different, so the kinks among them have different shape,  characteristic  speed, and behavior under collision. In some simple cases the dynamics of the soliton can be modeled heuristically via conservation laws, and certain observables---like its speed----can be computed in this way. A more general and complete approach would obviously be desirable.



The dynamical cactus, a simple phyllotactic  system depicted in \mbox{Figure \ref{fig1}}, consists of repulsive, massive objects holonomically constrained on rings rotating around a fixed axis: let $a=L/N$  be the distance between consecutive rings for a cylinder of length $L$ and radius $R$ containing $N$ objects;   $\theta_i$ is the angular coordinate of the $i^{th}$ particle, and $\omega_i=\theta_i-\theta_{i-1}$  the angular shift between consecutive rings. The total potential energy for unit length is then
\eq
V=\frac{1}{2L}\sum_{i \neq j} U(\theta_i-\theta_j)~,
\label{energy}
\en
where $U$ is any long ranged repulsive interaction that makes the sum extensive and well-behaved, such as a dipole-dipole or a  screened Coulomb. 

It has been shown that stable structures correspond to spiraling lattices~\cite{Nisoli}, where each ring is shifted from the previous by the same screw angle $\omega_i=\Omega$. The energy of spiraling structures as a function of $\Omega$ is reported in \mbox{Figure \ref{fig1}} for different values of $a/R$: as $a/R$ decreases, more and more commensurate spirals become energetically costly, independently of the repulsive interaction used; for every $j$ there is a value of $a/R$ low enough such that any commensurate spiral of screw angle $2\pi \!\ i/j$ with $i, j$ relatively prime, becomes a local maximum, as particles facing after $j$ rings becomes  nearest  neighbors in the real space. The minima also become more nearly degenerate as the density increases, since for angles incommensurate to $\pi$ each particle is embedded in a nearly uniform, incommensurately smeared background formed by the other particles~\cite{Nisoli}. 

One-dimensional degenerate systems are entropically unstable against domain wall formation~\cite{Lubensky}. Kinks between stable spirals were found numerically and experimentally~\cite{Nisoli}. Numerically it was proved that these kinks can propagate along the axis of the cylinder. Experimentally, the dynamical cactus was observed  expelling a higher energy domain by propagating its kink.  

As discussed in Ref.~\cite{Nisoli} such axial motion of a kink between two domains of different helical angles confronts a dilemma: helical phase is unwound from one domain at a different rate than it is wound up by the other. Numerical simulations show that the moving domain wall solves this problem by placing adjacent domains into relative rotation (see supplementary materials in Ref.~\cite{Nisoli} for an animation). 

Detailed dynamical simulations reveal  a complex zoology of solitons, along with an extraordinarily rich phenomenology: kinks of different species merge, decay, change identity upon collision, and decompose at high temperature into a sea of constituent lattice particles.

\section{Continuum Model.}
\subsection{Lagrangian.}
To describe domain walls between degenerate structures of the dynamical cactus we propose a minimal continuum model. First in the continuum limit we promote the  site index $i$ to a continuum variable: $i a\rightarrow z$, $\theta_i \rightarrow \theta(z)$, the screw angle is now $\omega_i \rightarrow \omega(z) \equiv a\d_z \theta$ (later we will just assume  $a=1$).  A spiral corresponds to $\omega(z)=\d_z \theta(z)=$constant. A kink corresponds, in the $\omega$, $z$ diagram, to a transition between two constant values. 

We seek a model that is local and  returns the right energy when the system is in a stable configuration.  To express the cost in bridging two stable domains we introduce a generalized rigidity $\kappa$.  We thus propose the  (linear density  of) Lagrangian  as
\eq
{\cal L} = \frac{1}{2} I \dot \theta ^2-V(\d_z \theta)- \frac{\kappa}{2} \(\d^2_z \theta\)^2.
\label{lag}
\en
The first term is the kinetic energy ($I \equiv {\mathcal I} a^{-1}$, ${\mathcal I}$ being the moment of inertia of a ring). The second accounts locally for the potential energy,  so that the hamiltonian from~(\ref{lag}) returns  the right energy for a static (or uniformly rotating) stable spiral like those observed experimentally. The last term, the lowest order correction allowed both in the field $\theta$ and in its derivatives, conveys the rigidity toward spacial variations of the screw angle and  ensures that a kink always corresponds to a positive energy excitation. 

The Euler-Lagrangia equation of motion follows directly from \mbox{Eq.~(\ref{lag})} as
\eq
I \d_t^2 \theta = \d_z V'\(\d_z \theta\)-\kappa ~ \d_z^4 \theta 
\label{motion}
\en
and corresponds to the conservation of the density angular momentum $I\d_t \theta$, where the density of current of angular momentum is $V'\(\d_z \theta\)-\kappa ~ \d_z^3 \theta $ ($V'$ is the derivative of $V$ with respect to its argument and represents the local torque).
%

%
%

In the following we will prove that, whatever the precise form of  the actual potential $V$, and  the geometric parameters of the problem, there exist  solitonic solutions of Eq.~(\ref{motion}) between local minima of $V$. We will  show how to derive many qualitative and quantitative results  by reducing the problem to an equivalent one-dimensional newtonian equation, as customary for these kind of problems~\cite{Kleinert}, yet with a significant twist.

\begin{figure}[t!!]
\center
\includegraphics[width=3.2 in]{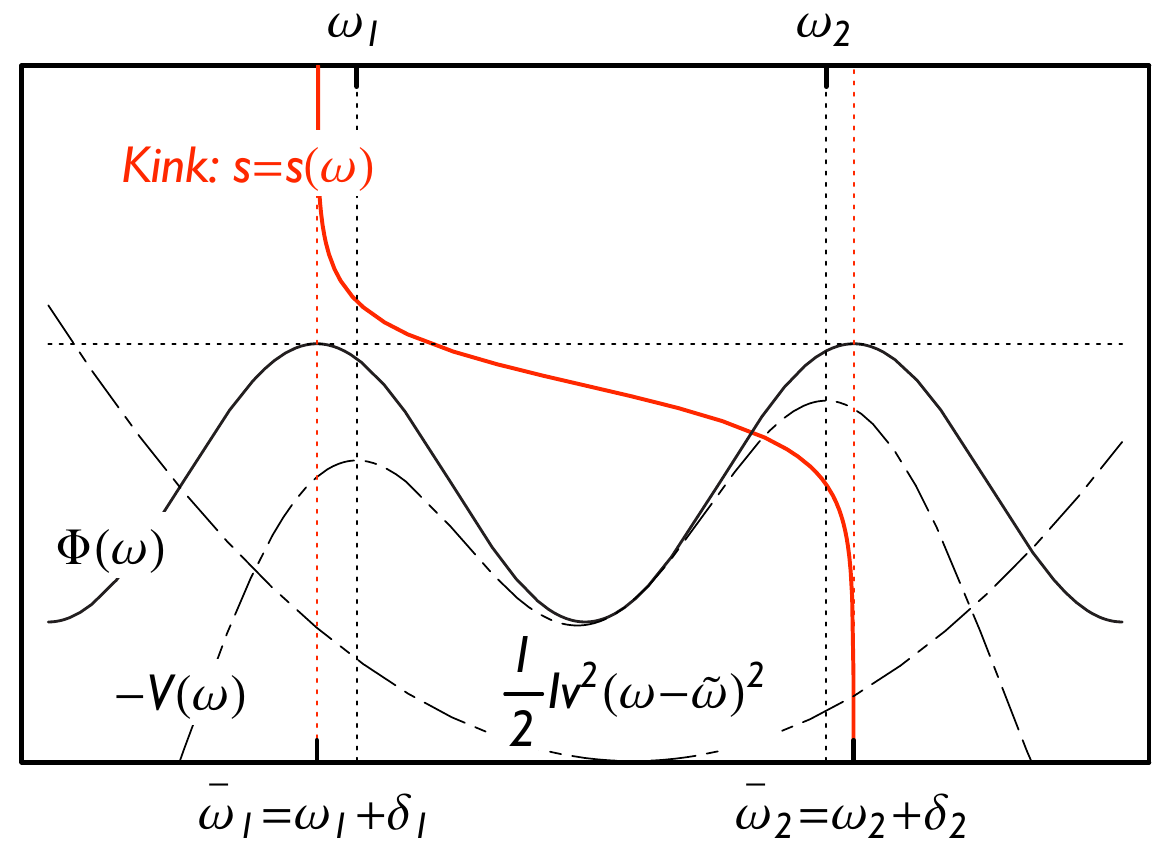}
\caption{(Color online). The equivalent newtonian picture, standard for Klein-Gordon-like solitons or for tunneling problems~\cite{Kleinert}:  a soliton is a  solution of Eq.~(\ref{Newton}) that corresponds to a trajectory $\omega(s)$ (red solid line [light gray]) of a point particle that starts and ends in two equally valued  maxima (horizontal dotted red lines [light gray]) of the equivalent potential energy $\Phi\(\omega\)=\frac{1}{2} v^2 I ( \omega-\tilde \omega)^2 -V\(\omega\)$ (black solid line).  As the kinetic energy must be zero on those maxima, it takes the newtonian trajectory an infinite amount of time $s$  (i.e. ``time'' in the newtonian picture,  space in the soliton picture) to reach the ``top of the hill'' with asymptotic speed equal to zero~\cite{Kleinert}. Unlike the Kline-Gordon-like case, the presence of the quadratic term  $\frac{1}{2} v^2 I ( \omega-\tilde \omega)^2$ (dashed line), which accounts for the kinetic energy of the domains, slightly shifts the  asymptotic value of the kinks (red dotted lines) away from stable static structures $\omega_1$, $\omega_2$ (black dotted lines), the minima of the  real potential energy $V\(\omega\)$ (dashed line). Instead the soliton connects spirals of angle $\bar \omega_1=\omega_1+\delta_1$, $\bar \omega_2=\omega_2+\delta_2$, which in general are not stable structures of $V$. The extra quadratic term also allows solitons among non-degenerate configurations, and brings the speed of the soliton $v$ into the picture. (Everything is in arbitrary units. The soliton is obtained by numerical integration of the Eq.~(\ref{Newton}) using the energy profile portrayed in the figure.)}
\label{fig2}
\end{figure}

\subsection{Equivalent newtonian picture.}

 We seek a uniformly translating solution of  Eq.~(\ref{motion})  
\eq
\theta=\theta (z-v t) + w t,
\label{sol}
\en
where the constant $w$ accounts for  angular rotation invariance, reflects the underlying transverse structure in our system (which is lacking in Llein-gordon-like systems), and is the source of much of the new physics.
By defining  $s=z-vt$, we find $\d^2_t \theta = v^2 \ud^2 \theta /\ud {s}^2$, which leads to the {\it equivalent newtonian equation}~\cite{Kleinert}
\eq
\kappa \frac{\ud^2 \omega}{\ud s^2}  = - \Phi' \(\omega\)
\label{Newton}
\en
whose {\it equivalent potential}  $\Phi$  is given by
\eq
\Phi\(\omega\)=\frac{1}{2} v^2 I ( \omega-\tilde \omega)^2 -V\(\omega\).
\label{equiv}
\en
$\tilde\omega$ is a constant that is sometimes useful to redefine as  $\tilde \omega\equiv \tau/(Iv^2)$ in terms of $\tau$, which has the dimension of a torque. We will find that in studying static solitons it is easier to use $\tau$, while for dynamical ones $\tilde \omega$ is preferable.

In  the equivalent newtonian picture a soliton corresponds  to a trajectory $\omega(s)$ of a point  particle that starts with zero kinetic energy on the top of a maximum of $\Phi(\omega)$ and, in an infinite amount of ``time'' $s$, reaches the neighboring maximum of equal height, as detailed in Figure~\ref{fig2}. A similar treatment applies to Klein-Gordon-like solitons and tunneling problems~\cite{Kleinert,Lubensky}. Yet the presence of an extra quadratic term translates in completely different physics and accounts for the dynamics of the domains. 

\subsection{Shift from stability.}

Because of the quadratic term $\frac{1}{2} v^2 I ( \omega-\tilde \omega)^2 $ in the expression of $\Phi$, and unlike the Klein-Gordon-like soliton, our soliton in general does not connect two stable structures $ \omega_1$,  $\omega_2$, local minima of $V$; rather it goes from $\bar \omega_1=\omega_1+\delta_1$ to $\bar \omega_2=\omega_2+\delta_2$, bridging spirals whose screw angle is slightly shifted away from stability,  the shift increasing with the speed of the kink (Figure~\ref{fig2}).

We find then a most interesting result: {\it a moving soliton cannot connect two stable structures}. At most, it can bridge a stable spiral $\omega_1$ with one shifted away from equilibrium $\bar \omega_2=\omega_2+\delta_2$, if we choose  $\tilde \omega=\omega_1$, but only if $V(\omega_1)<V(\omega_2)$. 

These shifts away from stability were observed in dynamical simulations (integrating the full equation of motion for the discrete system of particles interacting via Eq.~\ref{energy}) in the form of  precursor waves propagating at the speed of sound in front of the soliton, as in Figure~\ref{fig3}. They correspond to the physically intuitive fact that, in order to move, the kink needs a torque to propel it. 

Clearly, static solitons ($v=0$) can  bridge stable spirals if these are degenerate, i.e. , $V(\omega_1)=V(\omega_2)$.
On the other hand, if two domains are degenerate, and are connected by a moving soliton, they {\it both} need to be shifted. 

\subsection{Boundary conditions.}

The conditions of existence of a soliton between the asymptotic domains  $\bar \omega_1$, $\bar \omega_2$,
are
\begin{equation}
\left\{ \begin{array}{l} 
 \!\  \Phi'(\bar \omega_1) = \Phi'(\bar \omega_2)=0 \\
 \!\ \Phi(\bar \omega_1) = \Phi(\bar \omega_2).
\end{array} \right .
\label{general}
\end{equation}
Now, from Eq.~(\ref{sol}) we obtain the expression of  the angular velocity for a soliton of speed $v$,
\eq
\dot \theta(s)=-v \!\ \omega(s)+w,
\label{ang}
\en
which implies that in general, as the soliton travels along the axis, {\it one or both of its two domains must be set into rotation, and  at different angular velocities}. As mentioned, this rotation of the domains is  indeed observed in dynamical simulations~\cite{Nisoli} as the mechanism with which the soliton unwinds and rewinds spirals of different gaining angles. 

Unlike sine-Gordon-like one-dimensional topological solitons, which separate essentially equivalent  static domains and can travel at any subsonic speed~\cite{Lubensky}, we see that phyllotactic domain walls instead separate regions of different dynamics: energy and angular momentum {\it flow through} the topological soliton as it moves, rather than being concentrated in it; as mentioned above, this was used to show heuristically that its speed $v$ is tightly controlled by energy-momentum conservation, phase matching at the interface and boundary conditions~\cite{Nisoli}. We will show how  a more precise version of those heuristic formul\ae  ~can be deduced within the framework of our continuum model, and in a more general fashion. 

The angular speed of rotation of the domains depends on the parameter $w$, which along with $\tilde \omega$ is a constant to be determined. On the other hand we can show that energy conservation constrains  $w$ to $\tilde \omega$. In fact, when the two domains are shifted from equilibrium, and thus subjected to a torque, energy flows through the boundaries of the system as the domains rotate.
By imposing  no energy  accumulation in the kink we fix  $w$ in Eq.~(\ref{sol}) in the following way: the power entering the system at the asymptotic boundaries can easily be deduced  via the Noether theorem, to  be 
\eq
j_{\pm \infty}= - \lim_{s\rightarrow \pm \infty}V'(\omega(s))\dot \theta(s) ,
\en
not surprisingly, the torque times the angular speed. By requiring $j_{+\infty}$=$j_{-\infty}$ an equation for $w$, $\tilde \omega$ is found
\eq
w=v(\bar \omega_1+\bar \omega_2-\tilde \omega)
\label{w}
\en
which will come in  hand in many practical cases. 

\section{Zoology.}

\subsection{Static Kinks.} 

Let us explore static kinks first. Equations~(\ref{general})  reduce to
\begin{equation}
\left\{ \begin{array}{l} 
 \!\ \tau = V'(\bar \omega_1)=V'(\bar \omega_2) \\
 \!\ \tau(\bar \omega_1-\bar \omega_2)= V(\bar \omega_1)-V(\bar \omega_2).
\end{array} \right.
\label{static}
\end{equation}
The first of the two Equations~(\ref{static}) tell us that $\tau$ is a torque applied at the asymptotic boundaries of the system, while the second shows that no torque is necessary if the topological soliton connects degenerate structures, as we anticipated. 

Unlike topological solitons of the  Klein-Gordon class, here  {\it static kinks are allowed between non-degenerate domains, through an applied torque} that shifts the two structures out of the minima of $V$.  

\subsection{Solitons moving between {\it non} degenerate domains.}

Unlike the Klein-Gordon-like Lagrangian, our Lagrangian is not invariant under Poincar\'e group. Our  traveling soliton is not just  a boost of the static one, as it is clear from Figure~\ref{fig2}. In particular the shifts of the domains from stability increase with the speed $v$. By expanding $V\(\omega\)$ around its local minima and using \mbox{Eq.~(\ref{equiv})} one finds easily that small shifts are proportional to the square of the speed of the soliton, or  $\delta_1,~\delta_2 \propto v^2$.

\begin{figure}[t!!]
\center
\includegraphics[width=3.1 in]{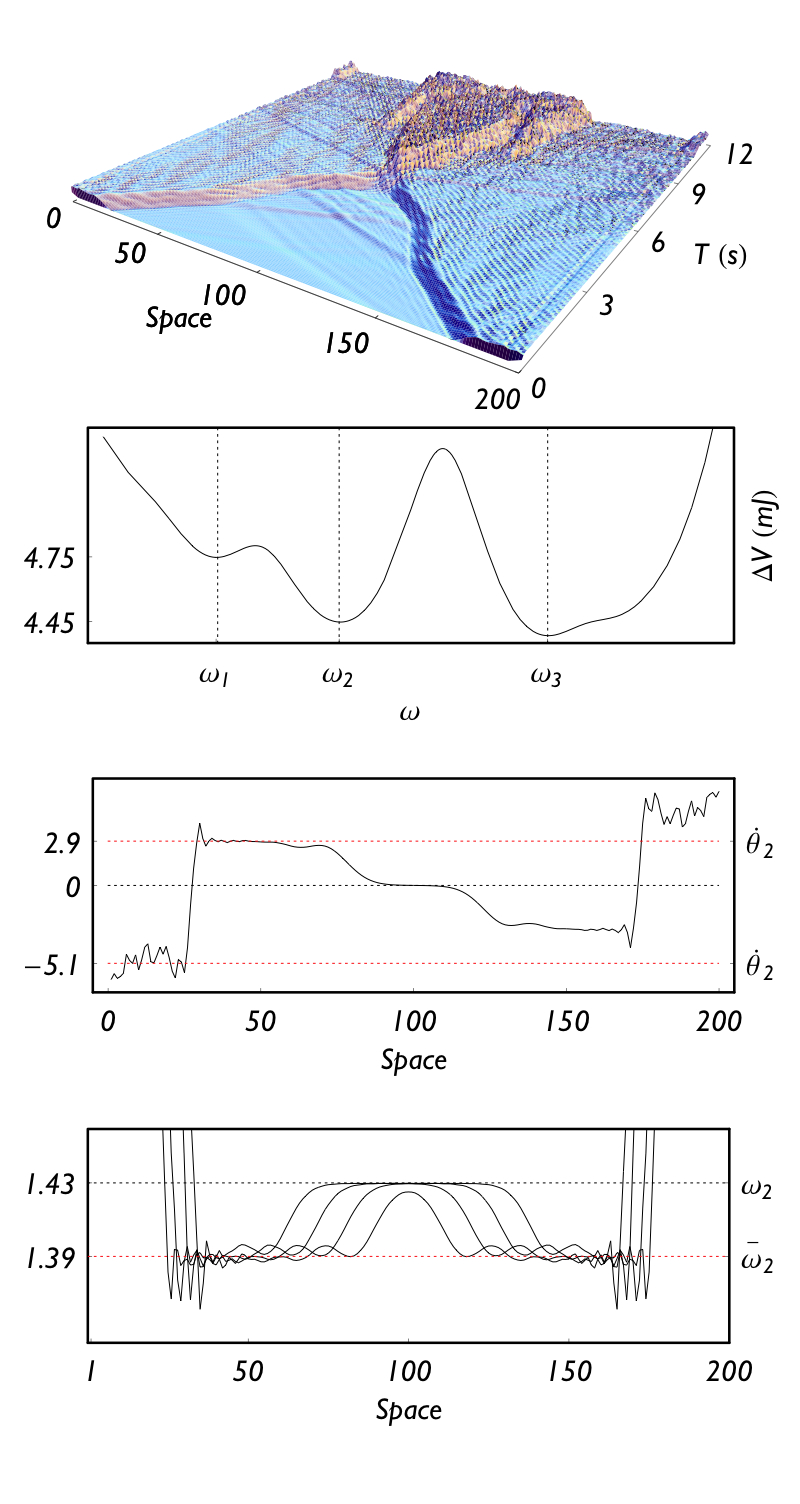}
\caption{(Color online). Dynamical simulation for the collision/conversion of two solitons, emitted from free boundaries of a dynamical cactus. Top panel: screw angle  {\it vs.} space (in number of rings) and time (in seconds); note the elastic wave preceding the soliton. Second panel: energy per particle (mJ) {\it vs.} screw angle (rad) for our system ($\omega_1=1.79$ rad, $\omega_2=1.43$ rad, $\omega_3=2.40$ rad); the two kinks  -- before collision -- connects a high (inner: $\omega_1$) to a low (outer: $\omega_2$)  energy domain; after collision a low (inner: $\omega_3$) to a high (outer: $\omega_2$).  Third panel: angular speed ($s^{-1}$) {\it vs.} space at a given time: the speed of the soliton can be extracted as $v=\Delta \dot \theta/\Delta \omega=22.1~s^{-1}$ (predicted $23.4 ~ s^{-1}$). Fourth Panel: the precursor; plot of the screw angle {\it vs.} space at different times while the soliton and its preceding wave advance.  The amplitude $\delta_1$ of the precursor is predicted via Eq.~\ref{d2} as $\delta_1 =\omega_1- \bar \omega_1=-0.043 ~\mathrm{rad}$, in excellent agreement with numerical observations.  The simulation uses the experimental density and magnetic  interaction as in Ref.~\cite{Nisoli}. (See supplementary materials in Ref.~\cite{Nisoli} for an animation.)}
\label{fig3}
\end{figure}

Let us consider a particular case of practical importance. 
Kinks between non-degenerate structures can sometimes be found in the experimental settings, because of static friction, after having annealed mechanically the cactus with free boundaries. If a small perturbation is given, enough to overcome that friction, the  dynamical cactus will expel the  higher energy domains by propagating the kink along the axis~\cite{Nisoli}. Figure~\ref{fig3} reports a dynamical simulation of this case obtained by integrating the full equation of motion for the discrete system of particles whose interaction is given in Eq.~(\ref{energy}).

Let us show how our continuum model can reproduce a soliton between non-degenerate minima and provide quantitative predictions. Free boundaries imply  $j_{\pm \infty}=0$: since $v\ne0$,  $V'$ cannot be zero on both boundaries, as we saw before, nor can the angular speed of rotation of the domains   be. The only possible free boundaries solution is one that has $
V'=0$ on one side and $\d_t \theta=0$ on the other. Hence the soliton must connect, say,   $\bar \omega_2=\omega_2$ to $\bar \omega_1=\omega_1 +\delta_1$, with $V(\omega_2)<V(\omega_1)$,  which fixes  $\tilde \omega=\omega_2=\bar \omega_2$. Equations~(\ref{general}) then fix the speed of the soliton as 
\eq
v_k^2=\frac{2 \Delta \bar V}{I \Delta \bar \omega^2}
\label{v}
\en
with $\Delta \bar V=V( \bar \omega_1) -V(\bar \omega_2)$, $\Delta \bar \omega=\bar \omega_2-\bar \omega_1$ (in this particular case $\omega_2=\bar \omega_2$). Eq.~(\ref{v}) corrects an earlier formula found heuristically via energy conservation, which neglected the shifts~\cite{Nisoli}. From the angular velocities of rotation of the domains in the dynamical simulation (\mbox{Figure \ref{fig3}}  third panel), a speed of $22.1 ~ \mathrm{s^{-1}}$  can be computed and compared with the value  $23.4 ~ \mathrm{ s^{-1}}$ predicted by \mbox{Eq.~(\ref{v})}: the small discrepancy is likely due to energy dissipation into phonons -- a known effect for solitons in discrete systems. 

From Eq.~(\ref{w}) we find  $w=v \!\ \bar \omega_2$ which together with Eq.~(\ref{ang}) implies that the region of lower energy rotates uniformly, while that of higher energy, which is shifted, remains still: since there is no energy flow through the boundaries, the kink  transforms the potential energy difference between the two domains into the kinetic energy of rotation -- and vice versa, depending on its direction of propagation. (Only propagation toward lower potential energies were observed experimentally~\cite{Nisoli}.) 

\begin{figure}[t!!]
\center
\includegraphics[width=3.5 in]{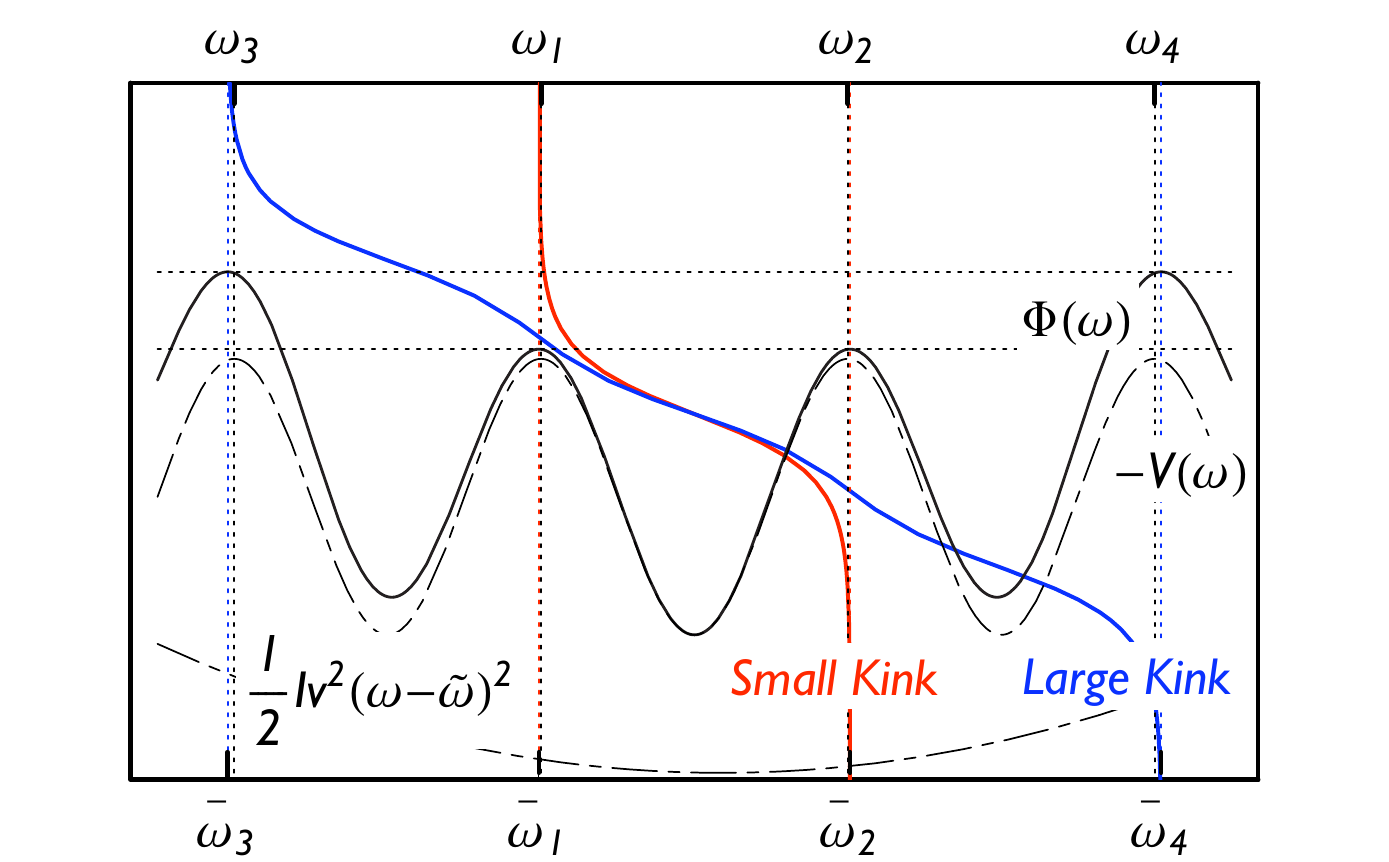}
\caption{(Color online). Equivalent newtonian diagram (see caption of Figure~\ref{fig2} for explanation) for symmetric solitons between degenerate spirals. As the quadratic term raises all the maxima of $-V\(\omega\)$, solitons with the same speed, but connecting different domain are possible under different applied torque. The smallest kink (solid red line [light gray]) and a larger kink (solid blue line [dark gray]) are shown.}
\label{fig4}
\end{figure}

By expanding  $V$ around $\omega_1$, $V\simeq \frac{1}{2}Ic_1^2(\omega-\omega_1)^2$, where $c_1$ is the speed of sound in the spiral of angle $\omega_1$, we obtain an approximate expression for the shift of the screw angle in the region of higher energy:
\eq
\frac{\delta_1}{\Delta \omega}= \frac{v^2/c_1^2}{1-v^2/c_1^2},
\label{d2}
\en
that clearly holds for $v\ll c_1$. As the system is prepared in a stable, non-shifted configuration of angle $\omega_1$, a precursor in front of the soliton propagates to accommodate  the shift $\delta_1$, as it was seen yet not understood in the dynamical simulation reported in \mbox{Figure \ref{fig3}}~\cite{Nisoli}.
Eq.~(\ref{d2}) applied to the experimentally measured $V(\omega)$ employed in the simulation, predicts a shift  $\delta_1=-0.043 ~\mathrm{rad}$, which fits the results of the simulation well  (\mbox{Figure \ref{fig3}}, last panel). 

We will not deal here with collisions between solitons. Yet, even without  knowledge of the precise form of $V$, a few considerations can be made on asymptotic states of the collision depicted in Figure~\ref{fig3}:  the domains beyond the approaching  pair of kink anti-kink rotate  with opposite angular velocity $\dot \theta=v \Delta \omega$, $v$ given by \mbox{Eq.~(\ref{v})}. After collision $v $ changes sign and so must $\Delta \omega$, as the infinitely long domains cannot invert angular velocity instantaneously: the emerging asymptotic configuration  must be that of a pair of different kink anti-kink, connecting the old asymptotic domain with a different nearest neighbor, hence the collision metamorphosis already discussed in Ref.~\cite{Nisoli}.  

\subsection{Solitons moving  between degenerate domains.}

To investigate the degenerate case, typical of dynamical phyllotaxis of high $R/a$ ratios (Figure~\ref{fig1}), we  consider for simplicity the potential $V$ of \mbox{Figure \ref{fig4}}; degeneracy requires now $\tilde \omega = (\omega_1 +\omega_2)/2$, and if the potential is symmetric as in the figure, both boundaries are equally shifted in opposite directions and subjected to a torque of opposite sign and equal intensity (from Equations~(\ref{general})) 
\eq
|\tau|=\frac{I}{2}  v^2 (\bar \omega_2-\bar \omega_1).
\en

Exactly as in the case of the boosted Klein-Gordon-like one, our symmetric soliton becomes shorter as $v$ increases (\mbox{Figure \ref{fig2}}), although via a completely different mechanism. In the  Klein-Gordon-like  soliton shortening is a consequence of relativistic contraction of the Poincar\'e group. Instead, in our case, $v$ raises the difference between the maxima  of $\Phi$ and the minimum among them. That means higher kinetic energy in the "valley" of $\Phi$ for the equivalent trajectory, and thus higher values of $d\omega/d s$, hence a shorter soliton.
 
\begin{figure}[b!!]
\center
\includegraphics[width=3.2 in]{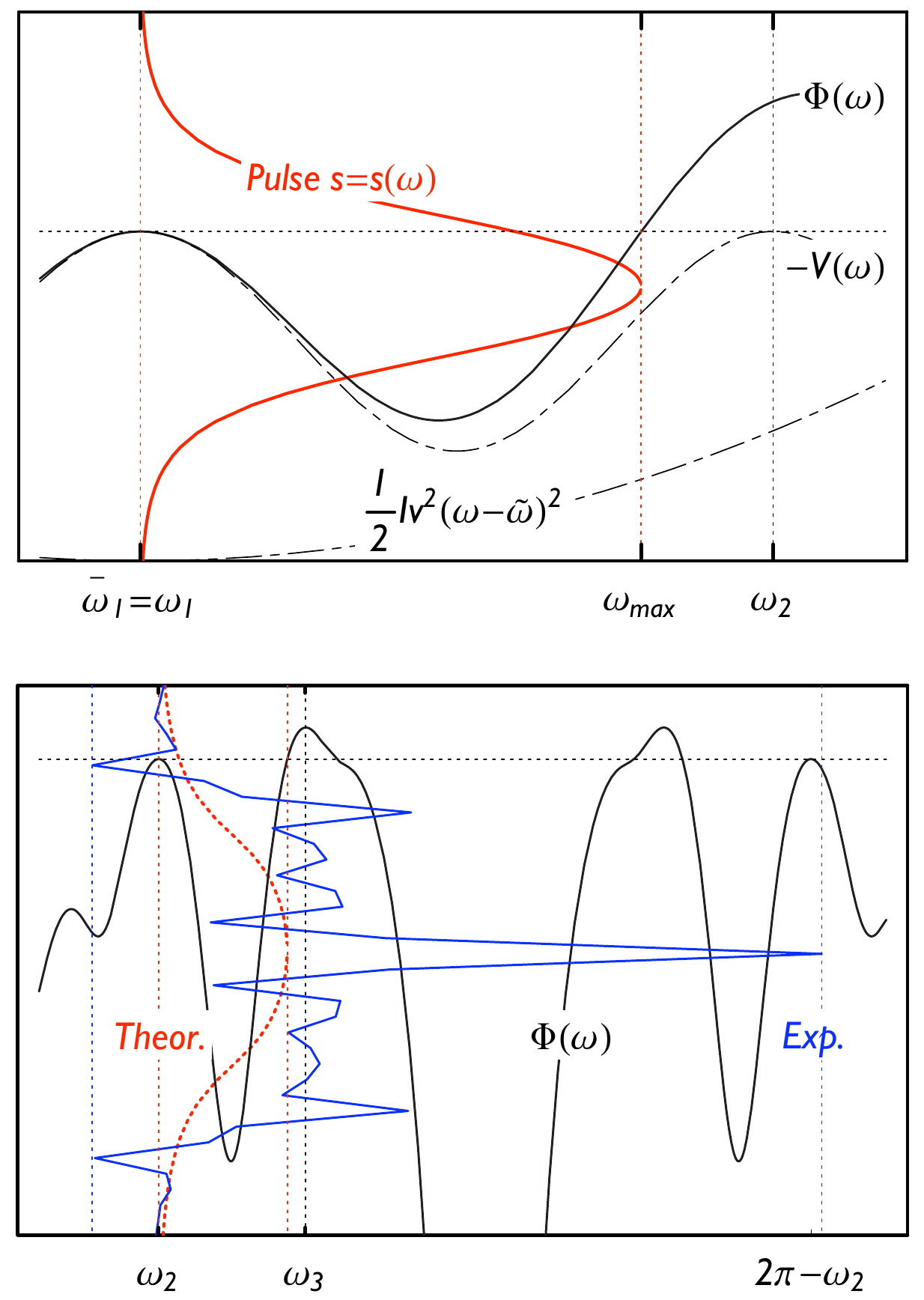}
\caption{(Color online). Equivalent newtonian diagram (see caption of Figure~\ref{fig2} for explanation) for pulses. Top: a traveling pulse  among degenerate structures  is possible because of the non-zero speed $v$. Bottom: theoretical (red dashed [light gray])  and experimental (blue solid [dark grey]) frozen-in pulse on the measured energy curve of the dynamical cactus~\cite{Nisoli}.}
\label{fig5}
\end{figure}

Equations~(\ref{ang})-(\ref{w}) tell us that the two domains rotate in  opposite direction with angular velocity of the same intensity
\eq
|\dot \theta| =v (\bar \omega_2-\bar\omega_1). 
\en
In practice, the system acts like a mechanical inverter of rotation, that transmits a power $\tau \dot\theta=I v^3 \Delta \bar \omega^2/2$ along the tube. 

For any given velocity, many solitons between degenerate structures are allowed,  corresponding to different applied torques, as in \mbox{Figure \ref{fig4}}.

\subsection{Pulses: propagating and frozen-in.}

The equation of motion also predicts pulses, both dynamic and static (frozen-in). The top panel of Figure~\ref{fig5} shows the pulse soliton in the equivalent newtonian picture as a trajectory falling from a local maximum  of $\Phi$, say $\omega_1$,  and coming back to it without reaching the neighboring $\bar \omega_2$ because of a higher potential barrier.

Clearly pulses posse less inertia than kinks, since only the region occupied by the pulse rotates during propagation. Theory suggests that these pulses should be able to propagate with free boundaries and that no applied torque is necessary -- although  solutions corresponding to different boundary conditions, and hence applied torque, can be found as well, by choosing $\tilde \omega \ne \omega_1$. 

Let us consider pulses in stable structures and thus $\tilde \omega = \omega_1$. When  $v>c$,  the speed of sound in the spiral of angle $\omega_1$,  then $\omega_1$ becomes a minimum for $\Phi$ and we obtain a solution that oscillates around $\omega_1$: {\it the speed of sound is the upper limit for the velocity of the pulse soliton}. 
As $v$ decreases three things can happen.  

If $V(\omega_1)<V(\omega_2)$, then pulses exist for $v_k<v<c$,   $v_k$ given in \mbox{Eq.~(\ref{v})}. As $v$ approaches from above the critical  value $v_k$, the pulse  asymptotically stretches to a pair of  kink and anti-kink between domains  $\omega_1$ and $\bar \omega_2$, placed at infinite distance from each other, traveling at the same speed  in the same direction. 

If $V(\omega_1)=V(\omega_2)$,  the situation is the same as above, yet the critical value for asymptotic stretch into kink and anti-kink is $v=0$.

If $V(\omega_1)>V(\omega_2)$, we have a pulse soliton for $0\le v<v_k$, when the velocity is less than the critical value given by Eq.~(\ref{v}). As $v$ goes to zero the pulse freezes into a static one.

Hence, in absence  of  torque at the boundaries, our theory predicts that {\it static pulses
} can exist in every structure, except in the lowest energy one.

These frozen-in kinks  were indeed observed  experimentally (but not understood) in higher energy spirals,  perhaps arising from a  kink-antikink symmetric collision~\cite{NisoliPRB} at the interplay with friction. \mbox{Figure \ref{fig5}}, bottom, shows in solid blue line (dark gray) data from the experimental apparatus, along with our frozen-in soliton calculated numerically, dotted red line (light gray) for the  energy  $\Phi=V$ (solid black line) empirically measured in the experimental apparatus. 

\section{Axially Unconstrained case.}

We have since now considered only an axially constrained case where the repulsive particles were allowed to rotate around an angular coordinate, but not to translate along the axis. Non-local optimization via a structural genetic algorithm has shown that the more general case of axially unconstrained particles on a cylindrical surface reproduces the same fundamental statics of the axially constrained one, i.e. the same spiraling lattices~\cite{NisoliPRB}. Numerical simulations have also shown kinks  propagating  among the spirals~\cite{NisoliPRB}. The most significant difference with the axially constrained case is a drop in density in the region of the kink, to locally relieve the mismatch.
If these particles were charged, the drop in density would endow the soliton with a net charge: the phyllotactic soliton could function as a charge carrier.

In fact, Wigner crystals of electrons on large semiconducting tubes are candidate environments for  the phyllotactic soliton at nanoscale, as discussed in Ref.~\cite{Nisoli}. A crystal pinned by the corrugation potential and/or impurities will not slide along the tube under a weak enough external field.

\begin{figure}[t!!]
\center
\vspace{1mm}\includegraphics[width=3.1 in]{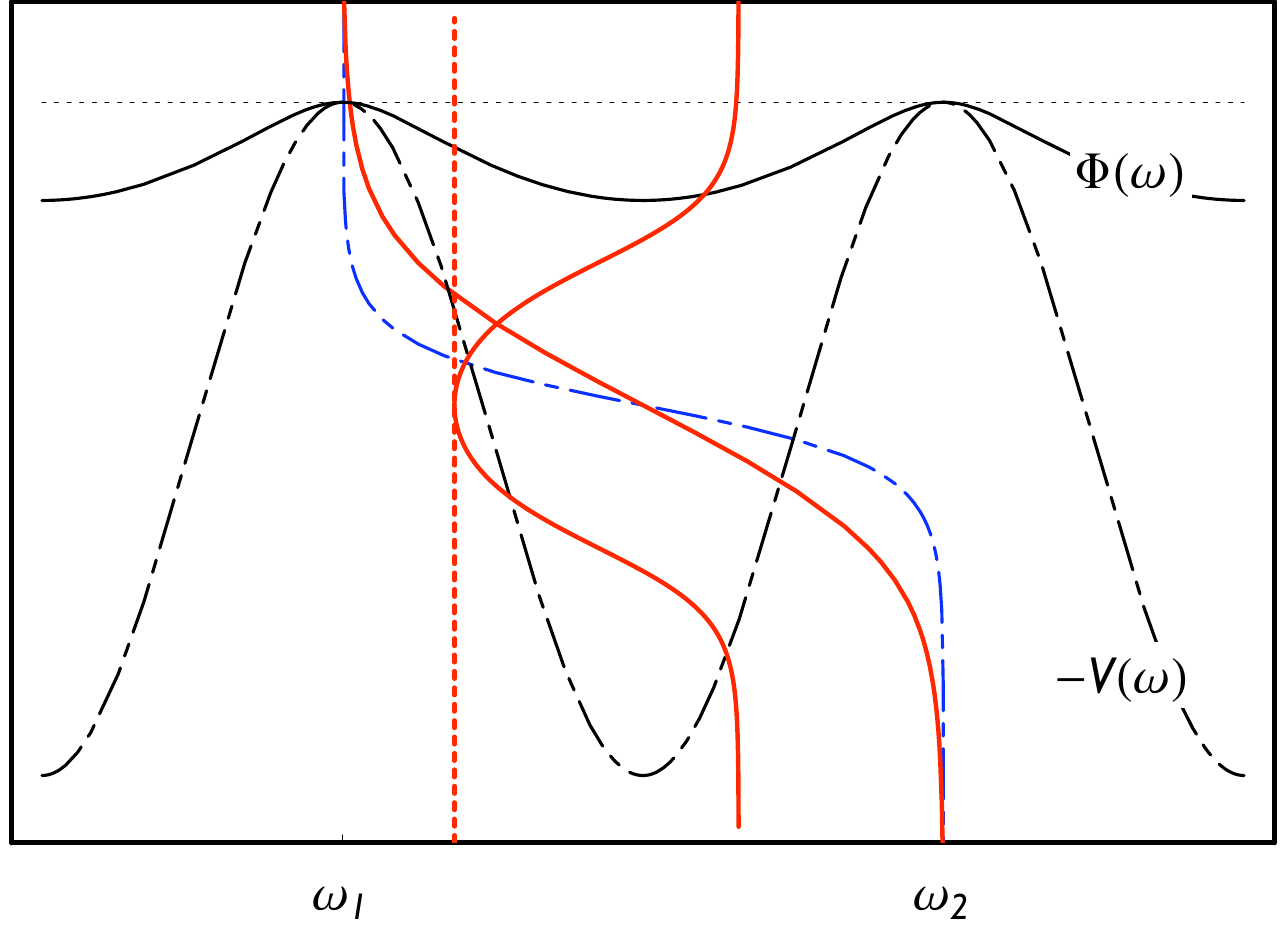}
\caption{(Color online). Equivalent newtonian diagram for both axially contrained and unconstrained soliton. In blue dashed (dark gray), the axially constrained soliton among the stable structures $\omega_1$, $\omega_2$. The two red solid curves (light gray) depict the soliton in the  axially unconstrained case--which is longer--and its corresponding drop in density $\lambda$.}
\label{fig6}
\end{figure}

A simple extension of the continuum model can easily incorporate variations in the linear density and allow investigation of  the possibility of charge transport by the phyllotactic soliton. We rescale  the  interaction as
\eq
V \rightarrow \(\frac{\lambda}{\lambda_o}\)^2 V
\en
where  $\lambda$  is the linear density in presence  of axial displacement, whereas $\lambda_o=N/L$.  Let $ \zeta=\ud z/a$ be the relative axial displacement: then, if $\d_z \zeta \ll1$ we have, $\lambda/ \lambda_o =1/(1+\d_z \zeta)$. 
The new Lagrangian reads
\eq
{\cal L} = \frac{1}{2} I \dot \theta ^2+\frac{1}{2} IR^{-2} \dot \zeta ^2-\frac{V(\d_z \theta)}{(1+\d_z \zeta)^2}- \frac{\kappa}{2} \(\d^2_z \theta\)^2
\label{lag2}
\en
To gain insight regarding the shape of the soliton we restrict ourself to the simpler static case with no applied torque. By variation of the lagrangian in Eq.~(\ref{lag2}) one obtains
\begin{equation}
\left\{ \begin{array}{l} 
 \!\ \frac{\ud^2 \omega}{\ud z^2}  = - \frac{V' \(\omega\)}{1+\d_z\zeta} \\
 \!\ \d_z \left[ \frac{2V\(\omega\)}{\(1+\d_z \zeta\)^3}\right]=0,
 \end{array} \right .
\end{equation} 
which, along with the normalization condition for $\lambda$, returns  the density of a static soliton (kink or pulse) as a function of $\omega(z)$
\eq
\frac{\lambda(z)}{\lambda_o}= \(\frac{V_{o}}{V\(\bar \omega\(z\)\)}\)^{2/3}.
\label{density}
\en
Here $V_{o}$ is the asymptotic  value of $V$ at the boundaries. The static kink $\bar \omega (z)$  can be found as the   the solution of the equivalent newtonian Eq.~(\ref{Newton}) with potential
\eq
\Phi\(\omega\)=-V_{o}^{2/3} V\(\omega\)^{1/3}.
\label{phinew}
\en
Now, $ V^{1/3}$ and $V$ have the same set of local minima, and the same ordering among 
their values: {\it the extension to axial displacements does not alter any of the conditions for existence of kink and pulse solitons}, at least for the static case. 

\mbox{Eq.~(\ref{density})} implies, as expected, a drop in density similar to a dark soliton in the region  of the kink, and thus, for a crystal of charges, a net charge. In particular, the higher the potential barrier between the two domains, the lower the density at the center of the kink.  Also, from Eq.~(\ref{phinew}) we can see that allowing an extra degree of freedom makes the kink longer. In practice the kink can take advantage of density reduction to relax the potential barrier between the two domains and can thus be longer. 
All this is detailed in Figure~\ref{fig6}, where the soliton for both the axially constrained and unconstrained case (along its variation in density) is reported for the same interaction among particles. Analogous considerations apply to static pulses. The case of the axially unconstrained traveling soliton is more complicated and will not be treated here.
 

\section{Conclusion.}

We have introduced a minimal, local, continuum model for the phyllotactic soliton and shown that its predictions are in excellent agreement with numerical data, that it  provides a tool to calculate otherwise elusive quantities, like charge transport,  energy-momentum flow, speed of the soliton and angular velocities, and could be used in the future to develop its thermodynamics.

Our continuum model should be applicable to many different physical systems. As  discussed in Ref.~\cite{Nisoli},  dynamical phyllotaxis is purely geometrical in origin, and thus the rich phenomenology of the phyllotactic soliton could appear across nearly every field of physics. Indeed phyllotactic domain walls have already been seen, but not recognized, in simulations~\cite{Schiffer} of cooled ion beams~\cite{Pallas} where the system self-organizes into concentric cylindrical shells. Colloidal particles on a cylindrical substrate provide a highly damped version~\cite{mesophase}, and polystyrene particles in air (as used to investigate~\cite{choi} the KTHNY theory of 2D melting~\cite{strandburg}) have reasonably low damping and long-range interaction. 

Yet it should be noticed our model has a range of application much wider than pure dynamical phyllotaxis. Domain walls in any spiraling system whose energy depends on the screw angle of the spirals, which manifest different stable spiraling structures,  and where a generalized rigidity can be reasonably introduced to describe transitions between different spirals, should be described by such a formalism.  Kinks in spiraling proteins and in particular transitions between A and B-DNA might be approachable this way.

The author would like to thank Vincent Crespi and Paul Lammert (Penn State University, University Park) for useful discussions, and  Ryan Kalas and Nicole Jeffery (Los Alamos National Laboratory) for helping with the manuscript.

This work was carried out under the auspices of the National Nuclear Security Administration of the U.S. Department of Energy at Los Alamos National Laboratory under Contract No. DE-AC52-06NA25396.


\begin{thebibliography}{}

\bibitem{solitons} M.~Remoissenet {\it Waves Called Solitons} (Springer-Verlag, Berlin, 1999)

\bibitem{Scott69} A.~C.~Scott Am. J. Phys. {\bf 37}, 52 (1969)

\bibitem{Nakajima} K.~Nakajima T.~Yamashita, Y.~Onodera Phys. Rev. B {\bf 45} 3141 (1974)

\bibitem{Ustinov} A.~V.~Ustinov  Physica  D {\bf 123} 315 (1998)

\bibitem{Kleinert} H.~Kleinert {\it Path Integrals} (World Scientific, Singapore, 1995)

\bibitem{Actor} A.~Actor Rev. Mod. Phys. {\bf 51} 461 (1979)

\bibitem{solitons2} N.~Manton, P.~Sutcliffe  {\it Topological Solitons} (Cambridge, 2004)

\bibitem{Nisoli} C.~Nisoli, N.~M.~Gabor, P.~E.~Lammert, J.~D.~Maynard, and V.~H.~ Crespi Phys. Rev. Lett. {\bf 102}, 186103 (2009).

\bibitem{Levitov} S.~L.~Levitov Phys. Rev. Lett. {\bf 66}, 224 (1991),  EuroPhys. Lett. {\bf 14}, 533 (1991). 

\bibitem{his1}A.~W.~ Thompson, {\em On Growth and Form,} Cambridge
  Univ. Press, NY, 1959 (first edition 1917).  R.~V.~Jean. {\em
    Phyllotaxis: a Systemic Study in Plant Morphogenesis ,} Cambridge
  Univ. Press, Cambridge 1994. 

\bibitem{his2}I.~ Adler, D~. Barabe,  and R.~V.~Jean,  Ann. Bot. London {\bf80}, 231-244 (1997)

\bibitem{polypeptide} A.~Frey-Wyssling  Nature {\bf 173}, 596 (1954). R.~O.~Erickson Science {\bf 181}, 705 (1973).

\bibitem{Benard}  N.~Rivier, R.~Occelli, J.~Pantaloni and A.~Lissowski, J. Phys. {\bf 45}, 49 (1984).


\bibitem{Lubensky} P.~M.~Chaikin and T.~C.~Lubensky, {\it Principles of Condensed Matter Physics} Cambridge Univ. Press, NY, 2000. 

\bibitem{NisoliPRB} C. Nisoli {\it et al.} In preparation.

\bibitem{Schiffer} A.~Rahman and J.~P.~Schiffer,  Phys. Rev. Lett. {\bf 57}, 1133 (1986);  J.~P.~Schiffer Proceeding of the Particle Accelerator Conference {\bf 5} 3264 (1995).

\bibitem{Pallas} T.~Shatz,  U.~Schramm and  D.~Habs, Nature {\bf 412}, 717 (2001);  U.~Schramm, T.~Shatz and D.~Habs, Phys. Rev. E {\bf 66}, 036501 (2002).

\bibitem{mesophase} D.~J.~Pochan et al., 
Science {\bf 306}, 94 (2004).

\bibitem{choi} Y.~Choi, K.~Kim and H.~K.~Pak, Physica A {\bf 281}, 99 (2000). 

\bibitem{strandburg} K.~J.~Strandburg, Rev Mod Phys {\bf 60}, 161 (1988).

\end{thebibliography}
\end{document}